\documentclass[12pt]{article}
\usepackage{amsmath}
\usepackage{amssymb}
\usepackage{amsfonts}
\usepackage{graphics}
\usepackage [latin1]{inputenc}

\title{Rationally-extended radial harmonic oscillator in a position-dependent mass background}
\author{C. Quesne\thanks{E-mail address: christiane.quesne@ulb.be}\\ 
{\small\sl D\'epartement de Physique,  Universit\'e Libre de Bruxelles,} \\ 
{\small\sl Campus de la Plaine CP229, Boulevard~du Triomphe, B-1050 Brussels, Belgium}}
\date{ }
\begin{document}
\baselineskip=22pt plus 1pt minus 1pt
\maketitle

\begin{abstract}
We show that the radial harmonic oscillator problem in the position-dependent mass background of the type $m(\alpha;r) = (1+\alpha r^2)^{-2}$, $\alpha>0$, can be solved by using a point canonical transformation mapping the corresponding Schr\"odinger equation onto that of the P\"oschl-Teller I potential with constant mass. The radial harmonic oscillator problem with position-dependent mass is shown to exhibit a deformed shape invariance property in a deformed supersymmetric framework. The inverse point canonical transformation then provides some exactly-solvable rational extensions of the radial harmonic oscillator with position-dependent mass associated with $X_m$-Jacobi exceptional orthogonal polynomials of type I, II, or III. The extended potentials of type I and II are proved to display deformed shape invariance. The spectrum and wavefunctions of the radial harmonic oscillator potential and its extensions are shown to go over to well-known results when the deforming parameter $\alpha$ goes to zero.
\end{abstract}


%
%
\newpage
\section{Introduction}

It is well known that the concept of position-dependent mass (PDM) in quantum mechanical systems has a lot of applications in several fields, such as electronic properties of semi-conductors and quantum dots, quantum liquids, $^3$He clusters, metal clusters, and energy density many-body problems \cite{bastard, weisbuch, serra, harrison, barranco, geller, arias, puente, ring, bonatsos, willatzen, chamel}. Finding exact solutions of the Schr\"odinger equation containing a PDM is therefore very useful in such contexts.\par
%
%
It is also worth noting that the PDM presence in the Schr\"odinger equation may alternatively be seen as reflecting a curvature of the underlying space or a deformation of the canonical commutation relations \cite{cq04}.\par
%
%
Among the methods used to build exact solutions of a PDM Schr\"odinger equation, one of the most powerful is the point canonical transformation (PCT) applied to a constant-mass Schr\"odinger equation, whose solutions are well known \cite{bagchi04, cq09a}. There are many applications of this method, among which one may quote some recent ones \cite{cq21, cq22, cq23a}.\par
%
%
Another topic that has recently found a lot of attention is that of the construction of exactly-solvable rational extensions of well-known quantum potentials. Such a subject arose after the introduction of exceptional orthogonal polynomials (EOPs) \cite{gomez09}, which generalize the classical orthogonal polynomials of Hermite, Laguerre, and Jacobi \cite{olver} in the sense that they form orthogonal and complete polynomial sets although there are some gaps in the sequence of their degrees in contrast with classical orthogonal polynomials. It turned out that such exactly-solvable rationally-extended potentials are related to Darboux transformations in the context of shape invariance in supersymmetric quantum mechanics (SUSYQM) \cite{cq08, cq09b, odake09, gomez11, odake11}.\par
%
%
{}For generalized Schr\"odinger equations, there have been less attempts of building exactly-solvable rational potentials connected with EOP's, among which one may quote some extensions of the quantum oscillator and Kepler-Coulomb problems in curved space \cite{cq16}, those of an oscillator-shaped quantum well confined in a cavity between two infinite walls \cite{cq23b}, as well as some families of asymmetric parabolic potentials associated with different kinds of PDM \cite{yadav}.\par
%
%
The purpose of the present paper is threefold: first to solve the problem of a three-dimensional radial harmonic oscillator potential in the background of a $m(\alpha;r) = (1+\alpha r^2)^{-2}$ PDM with $\alpha>0$, second to build some rational extensions of such a potential connected with $X_m$-Jacobi EOP's, and third to relate the latter to the well-known rational extensions of the radial harmonic oscillator potential in a constant-mass background, connected with $X_m$-Laguerre EOP's, when $\alpha$ goes to zero.\par
%
%
For this aim, it will prove essential to establish the existence of a PCT relating the radial harmonic oscillator problem in the $m(\alpha; r)$ PDM background to the P\"oschl-Teller I  (PT I) potential problem in a constant-mass background.\par
%
%
The plan of the paper is as follows. In Section~2, the existence of such a PCT is established, shown to solve the radial harmonic oscillator problem in the $m(\alpha; r)$ background, and used to prove the deformed shape invariance of the latter. In Section~3, some rational extensions of the radial harmonic oscillator problem are then constructed and their properties are studied. Finally, Section~4 contains the conclusion.\par
%
%
\section{\boldmath The radial harmonic oscillator in a $m(\alpha; r) = (1+\alpha r^2)^{-2}$ PDM background}

It is well known that one of the problems of PDM Schr\"odinger equations comes from the momentum and mass operator noncommutativity and the resultant ordering ambiguity in the kinetic energy term. To cope with this difficulty, one may use the von Roos general two-parameter form of the kinetic energy operator \cite{vonroos}. In units wherein $\hbar = 2m_0 = 1$, this would lead to the radial Schr\"odinger equation
\begin{align}
  & \biggl[- \frac{1}{2}\biggl(m^{\xi}(\alpha;r) \frac{d}{dr} m^{\eta}(\alpha;r) \frac{d}{dr} m^{\zeta}(\alpha;r) +
      m^{\zeta}(\alpha;r) \frac{d}{dr} m^{\eta}(\alpha;r) \frac{d}{dr} m^{\xi}(\alpha;r) \biggr)
      + V_{\rm eff}(r) \biggr] \psi(r) \nonumber \\[0.2cm]
  & = E \psi(r), \qquad 0 < r < \infty,  \label{eq:vonroos}
\end{align}
where $m(\alpha;r)$ denotes the dimensionless form of the mass function $m_0 m(\alpha;r)$ and the von Roos ambiguity parameters $\xi$, $\eta$, $\zeta$ are constrained by the condition $\xi + \eta + \zeta = -1$.\par
%
%
In the present paper, we plan to adopt the Mustafa-Mazharimousavi ordering \cite{mustafa}, corresponding to $\xi = \zeta = -1/4$ and $\eta = - 1/2$, which is one of the two orderings that pass the de Souza Dutra and Almeida test \cite{desouza} as good orderings. Our starting radial PDM Schr\"odinger equation will therefore be
\begin{equation}
  \biggl(- m^{-\frac{1}{4}}(\alpha;r) \frac{d}{dr} m^{-\frac{1}{2}}(\alpha;r) \frac{d}{dr} m^{-\frac{1}{4}}(\alpha;r)
  + V(r)\biggr) \psi^{(\alpha)}_n(r) = E^{(\alpha)}_n \psi^{(\alpha)}_n(r).  \label{eq:MM}
\end{equation}
We note that the general von Roos equation (\ref{eq:vonroos}) would reduce to the Mustafa-Mazharimousavi form (\ref{eq:MM}) by taking 
\begin{equation}
  V_{\rm eff}(r) = V(r) + \frac{1}{2}\left(\xi+\zeta+\frac{1}{2}\right) \frac{m''}{m^2} - \left(
  \xi\zeta + \xi + \zeta + \frac{7}{16}\right) \frac{m'^2}{m^3},
\end{equation}
where a dash denotes a derivative with respect to $r$.\par
%
%
The present choice for $m(\alpha; r)$ and $V(r)$ is
\begin{equation}
  m(\alpha;r) = \frac{1}{(1+\alpha r^2)^2}, \qquad \alpha>0,
\end{equation}
and
\begin{equation}
  V(r) = V(r;L,\omega) = \frac{L(L+1)}{r^2} + \frac{1}{4} \omega^2 r^2, \qquad 0 < r < \infty,  \label{eq:V}
\end{equation}
where $L=0$, 1, 2, \ldots\ is the orbital angular momentum quantum number and $\omega$ is the oscillator frequency. With such choices of PDM and potential, the radial Schr\"odinger equation (\ref{eq:MM}) may also be written as
\begin{equation}
  [\pi_r^2 + V(r;L,\omega)] \psi^{(\alpha)}_n(r;L,\omega) = E^{(\alpha)}_n(L,\omega) \psi^{(\alpha)}_n(r;L,
  \omega),  \label{eq:def-SE}
\end{equation}
where $\pi_r$ denotes the deformed momentum operator
\begin{equation}
  \pi_r = - {\rm i} \sqrt{f(\alpha;r)} \frac{d}{dr} \sqrt{f(\alpha;r)}, \qquad f(\alpha;r) = 1 + \alpha r^2 \qquad (\alpha
  >0),  \label{eq:f}
\end{equation}
written in terms of a deforming function $f(\alpha;r)$.\par
%
%
In a first step, we plan to solve Eq.~(\ref{eq:def-SE}) by resorting to the PCT method.\par
%
%
\subsection{PCT method applied to the radial harmonic oscillator potential in the $m(\alpha;r)$ background}

A deformed Schr\"odinger equation of type (\ref{eq:def-SE}) can be mapped onto a constant-mass Schr\"odingert equation
\begin{equation}
  \left(- \frac{d^2}{du^2} + U(u)\right) \phi_n(u) = \epsilon_n \phi_n(u). \label{eq:CM-SE}
\end{equation}
by some changes of variable and of function \cite{bagchi04, cq09a}
\begin{align}
  & u(\alpha;r) = a v(\alpha;r) + b, \qquad v(\alpha;r) = \int^r \frac{dr'}{f(\alpha;r')}, \label{eq:PCT1}\\
  & \phi_n(u(\alpha;r)) \propto \sqrt{f(\alpha;r)}\, \psi^{(\alpha)}_n(r;L,\omega).  \label{eq:PCT2}
\end{align}
The two potentials and their corresponding energy eigenvalues are related by
\begin{equation}
  V(r;L,\omega) = a^2 U(u) + c, \qquad E^{(\alpha)}_n(L,\omega) = a^2 \epsilon_n + c.  \label{eq:PCT3}
\end{equation}
In (\ref{eq:PCT1}) and (\ref{eq:PCT3}), $a$, $b$, and $c$ denote three arbitrary real constants.\par
%
%
{}For the choice of $f(\alpha;r)$ made in (\ref{eq:f}), we obtain
\begin{equation}
  v(\alpha;r) = \frac{1}{\sqrt{\alpha}} \arctan (\sqrt{\alpha} r).
\end{equation}
Hence, on taking
\begin{equation}
  a = \sqrt{\alpha}, \qquad b=0,  \label{eq:ab}
\end{equation}
we get
\begin{equation}
  u(\alpha;r) = \arctan(\sqrt{\alpha} r) \qquad \text{or} \qquad r = \frac{1}{\sqrt{\alpha}} \tan u,  \label{eq:u-r}
\end{equation}
showing that $u$ varies in the interval $0<u<\frac{\pi}{2}$.\par
%
%
With such a choice of variable, the potential $V(r;L,\omega)$, defined in (\ref{eq:V}), becomes
\begin{align}
  V(r;L,\omega) &= \alpha L(L+1) \cot^2 u + \frac{\omega^2}{4\alpha} \tan^2 u \nonumber \\
  &= \alpha L(L+1) \csc^2 u + \frac{\omega^2}{4\alpha} \sec^2 u - \alpha L(L+1) - \frac{\omega^2}{4\alpha}.
\end{align}
On comparing such an equation with (\ref{eq:PCT3}), we obtain that the potential of the constant-mass Schr\"odinger equation (\ref{eq:CM-SE}) is nothing else than the PT I potential\footnote{Note that the PT I potential is also known as the trigonometric P\"oschl-Teller or Darboux-P\"oschl-Teller potential and that it is related to the Scarf I potential $U_S(\bar{u}) = [\bar{A}(\bar{A}-1) + \bar{B}^2] \sec^2 \bar{u} - \bar{B}(2\bar{A}-1) \sec \bar{u} \tan \bar{u}$, $-\pi/2 < \bar{u} < \pi/2$ by the changes of parameters and of variable $\bar{A} = (A+B)/2$, $\bar{B}=(-A+B)/2$, $\bar{u} = 2u-\pi/2$ \cite{cq12}.}
\begin{equation}
  U(u) = U(u;A,B) = A(A-1) \csc^2 u + B(B-1) \sec^2 u, \qquad 0 < u < \frac{\pi}{2},  \label{eq:PT I}
\end{equation}
with
\begin{equation}
  A = L+1, \qquad B = \frac{1}{2}\left(1 + \frac{\Delta}{\alpha}\right), \qquad \Delta = \sqrt{\omega^2 +
   \alpha^2},  \label{eq:AB}
\end{equation}
while the additional constant $c$ in (\ref{eq:PCT3}) is given by
\begin{equation}
  c = - \alpha L(L+1) - \frac{\omega2}{4\alpha}.  \label{eq:c}
\end{equation}
\par
%
%
The PT I potential (\ref{eq:PT I}) is known to have an infinite number of bound-state wavefunctions \cite{poschl, cooper}
\begin{align}
  & \phi_n(u;A,B) = {\cal N}^{(A,B)}_n  (\sin u)^A (\cos u)^B P_n^{(A-\frac{1}{2}, B-\frac{1}{2})} (\cos 2u), 
      \label{eq:phi}\\
  & {\cal N}^{(A,B)}_n = \left(\frac{2(A+B+2n) n! \Gamma(A+B+n)}{\Gamma(A+n+\frac{1}{2})
      \Gamma(B+n+\frac{1}{2})}\right)^{1/2},
\end{align}
written in terms of Jacobi polynomials and corresponding to the eigenvalues
\begin{equation}
  \epsilon_n(A,B) = (A+B+2n)^2, \qquad n=0, 1, 2, \ldots.  \label{eq:epsilon}
\end{equation}
\par
%
%
{}From Eqs.~(\ref{eq:PCT3}), (\ref{eq:ab}), (\ref{eq:AB}), (\ref{eq:c}), and (\ref{eq:epsilon}), it therefore turns out that the PDM equation (\ref{eq:def-SE}) eigenvalues are given by
\begin{align}
  E^{(\alpha)}_n(L,\omega) &= \alpha \left(2L+\frac{5}{2}\right) + \left(L+\frac{3}{2}\right)\Delta + 4\left[
       \alpha\left(L+\frac{3}{2}\right) + \frac{1}{2} \Delta\right]n + 4 \alpha n^2, \nonumber \\ 
  &\qquad n=0,1,2,\ldots,  \label{eq:E}
\end{align}
provided the corresponding wavefunctions are normalizable on $(0, \infty)$. This is indeed the case, the latter being given by
\begin{equation}
  \psi^{(\alpha)}_n(r;L,\omega) = \lambda [f(\alpha;r)]^{-1/2} \phi_n(u;A,B),  \label{eq:psi-phi}
\end{equation}
where $\lambda$ is the constant $\alpha^{1/4}$, arising from the change of normalization from $u\in \left(0, \frac{\pi}{2}\right)$ to $r \in (0, \infty)$. The result for $\psi^{(\alpha)}_n(r;L,\omega)$ reads
\begin{equation}
  \psi^{(\alpha)}_n(r;L,\omega) = {\cal N}^{(\alpha)}_{n,L,\omega} r^{L+1} [f(\alpha;r)]^{-\frac{1}{2}\left(L
  +\frac{5}{2}+\frac{\Delta}{2\alpha}\right)} P_n^{\left(L+\frac{1}{2}, \frac{\Delta}{2\alpha}\right)}(t), \qquad
  t = \frac{1-\alpha r^2}{1+\alpha r^2}, 
\end{equation}
with
\begin{equation}
  {\cal N}^{(\alpha)}_{n,L,\omega} = \left(\frac{2 \alpha^{L+\frac{3}{2}} n! \left(2n+L+\frac{3}{2}+\frac{\Delta}
  {2\alpha}\right) \Gamma\left(n+L+\frac{3}{2}+\frac{\Delta}{2\alpha}\right)}{\Gamma\left(n+L+\frac{3}{2}\right)
  \Gamma\left(n+1+\frac{\Delta}{2\alpha}\right)}\right)^{1/2},
\end{equation}
and $n=0$, 1, 2, \ldots.\par
%
%
It is worth observing that when $\alpha$ goes to zero, so that Eq.~(\ref{eq:def-SE}) gives back the constant-mass Schr\"odinger equation for the radial harmonic oscillator
\begin{equation}
  \left(- \frac{d^2}{dr^2} + \frac{L(L+1)}{r^2} + \frac{1}{4} \omega^2 r^2\right) \psi^{(0)}_n(r;L,\omega)
  =E^{(0)}_n(L,\omega)\, \psi^{(0)}_n(r;L,\omega),  \label{eq:SE}
\end{equation}
we get for $E^{(0)}_n(L,\omega)$ the usual linearly-spaced spectrum
\begin{equation}
  E^{(0)}_n(L,\omega) = \left(2n + L + \frac{3}{2}\right) \omega, \qquad n=0, 1, 2,
\end{equation}
with the well-known wavefunctions written in terms of Laguerre polynomials
\begin{equation}
\begin{split}
  &\psi^{(0)}_n(r;L,\omega) = {\cal N}^{(0)}_{n,L,\omega} r^{L+1} e^{-\frac{1}{4} \omega r^2}
      L^{\left(L+\frac{1}{2}\right)}_n\left(\frac{1}{2}\omega r^2\right), \\
  &{\cal N}^{(0)}_{n,L,\omega} = \left(\frac{\omega}{2}\right)^{\frac{1}{2}\left(L+\frac{3}{2}\right)}
      \left(\frac{2 n!}{\Gamma\left(n+L+\frac{3}{2}\right)}\right)^{1/2}. 
\end{split}
\end{equation}
Here, use is made of the relations
\begin{equation}
  \lim_{\alpha\to 0} f^{-\frac{1}{2}\left(L+\frac{5}{2}+\frac{\Delta}{2\alpha}\right)} = \lim_{\alpha\to 0}
  (1+\alpha r^2)^{-\frac{\omega}{4\alpha}} = e^{-\frac{1}{4} \omega r^2},   \label{eq:limit-1}
\end{equation}
and
\begin{align}
  \lim_{\alpha\to 0} P_n^{\left(L+\frac{1}{2}, \frac{\Delta}{2\alpha}\right)}(t) &= \lim_{\alpha\to 0}
      P_n^{\left(L+\frac{1}{2}, \frac{\omega}{2\alpha}\right)}(1 - 2\alpha r^2) = 
      \lim_{\beta \to 0} P_n^{\left(L+\frac{1}{2},\beta
      \right)}\left(1 - \frac{\omega r^2}{\beta}\right) \nonumber \\
  &= L_n ^{\left(L+\frac{1}{2}\right)}\left(\frac{1}{2}\omega r^2\right) \qquad \text{with $\beta = \frac{\omega}
      {2\alpha}$},  \label{eq:limit-2}
\end{align}
coming from Eqs.~(4.4.17) and (18.7.21) of Ref.~\cite{olver}, respectively.\par
%
%
In Fig.~1, the first few eigenvalues $E^{(\alpha)}_n(L,\omega)$ are plotted in terms of the deforming parameter $\alpha$ and the ground-state wavefunction $\psi^{(\alpha)}_0(r;L,\omega)$ is displayed for some $\alpha$ values in Fig.~2.\par
%
%
\begin{figure}
\begin{center}
\includegraphics{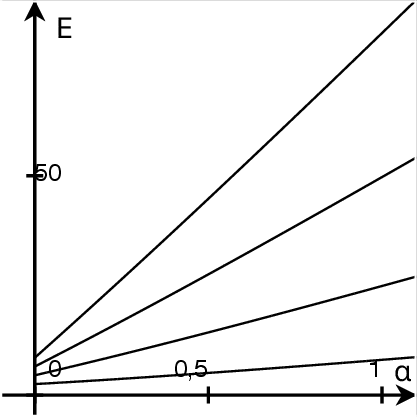}
\caption{Plots of $E^{(\alpha)}_n(L,\omega)$, defined in Eq.~(\ref{eq:E}), in terms of $\alpha$ for $n=0$, 1, 2, and 3. The parameter values are $L=\omega =1$.}
\end{center}
\end{figure}
\par
\begin{figure}
\begin{center}
\includegraphics{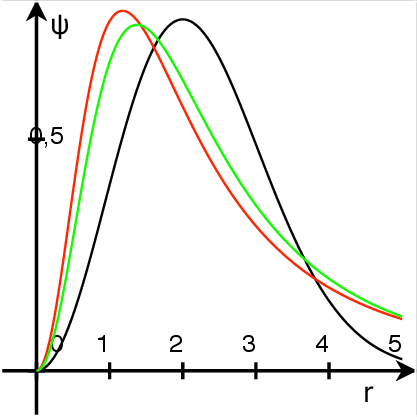}
\caption{Plots of the ground-state wavefunction $\psi^{(\alpha)}_0(r;L,\omega)$ in terms of $r$ for $\alpha = 1/\sqrt{3}$ (red line), $\alpha=1/(2\sqrt{2})$ (green line), and $\alpha=0$ (black line). The parameter values are $L=\omega=1$.}
\end{center}
\end{figure}
\par

%
%
\subsection{\boldmath Deformed shape invariance of the radial harmonic oscillator potential in the $m(\alpha;r)$ background}

It is well known that the PT I potential $U(u;A,B)$, defined in (\ref{eq:PT I}), is a shape invariant potential in the context of SUSYQM \cite{cooper}. As a result, the transformed potential $V(r;L,\omega)$ of Eq.~(\ref{eq:V}) is deformed shape invariant in the context of deformed SUSYQM \cite{bagchi05}.\par
%
%
This means that on taking the ground-state wavefunction $\psi^{(\alpha)}_0(r;L,\omega) = {\cal N}^{(\alpha)}_{0,L,\omega} r^{L+1} [f(\alpha;r)]^{-\frac{1}{2}\left(L+\frac{5}{2}+ \frac{\Delta}{2\alpha}\right)}$ of the deformed Schr\"odinger equation (\ref{eq:def-SE}) as the seed function, the corresponding superpotential
\begin{equation}
  W_{\alpha}(r;L,\omega) = - f \frac{d}{dr} \log \psi^{(\alpha)}_0(r;L,\omega) - \frac{1}{2} f',
\end{equation}
where a dash denotes a derivative with respect to $r$, can be written as
\begin{equation}
  W_{\alpha}(r;L,\omega) = - (L+1) \frac{f}{r} + \alpha\left(L + \frac{3}{2} + \frac{\Delta}{2\alpha}\right) r.
\end{equation}
It allows us to consider a pair of first-order differential operators
\begin{equation}
  A^{\pm}_{\alpha}(L,\omega) = \mp \sqrt{f} \frac{d}{dr} \sqrt{f} + W_{\alpha}(r;L,\omega),
\end{equation}
in terms of which the starting deformed Hamiltonian $H_0 = \pi_r^2 + V_0(r)$, $V_0(r) = V(r;L,\omega)$, can be written as
\begin{equation}
  H_0 = A^+_{\alpha}(L,\omega) A^-_{\alpha}(L,\omega) + E^{(\alpha)}_0(L,\omega).
\end{equation}
\par
%
%
The partner of $H_0$ is
\begin{align}
  H_1 &= A^-_{\alpha}(L,\omega) A^+_{\alpha}(L,\omega) + E^{(\alpha)}_0(L,\omega) = \pi_r^2 + V_1(r),
       \nonumber \\
  V_1(r) &= V_0(r) + 2f \frac{d}{dr} W_{\alpha}(r;L,\omega),
\end{align} 
where it turns out that
\begin{align}
  V_1(r) &= V_0(r;L+1,\omega') + \alpha \left(2L+3+\frac{\Delta}{\alpha}\right)  \nonumber \\
  &= \frac{(L+1)(L+2)}{r^2} + \frac{1}{4} \omega^{\prime 2}r^2 + \alpha\left(2L+3+ \frac{\Delta}{\alpha}\right),
\end{align}
with $\omega' = \sqrt{\omega^2+ 4\alpha^2+4\alpha\Delta}$ and $\Delta' = \sqrt{\omega^{\prime2}+\alpha^2} = \Delta+2\alpha$. Hence, the potential $V_0(r) = V(r;L,\omega)$ exhibits a deformed shape invariance property: up to some additive constant, its partner $V_1(r)$ is similar in shape and differs only in the parameters $L+1$, $\omega'$ that appear in it.\par
%
%
Such a property enables us to construct a hierarchy of Hamiltonians
\begin{equation}
  H_i = A^+_{\alpha}(L+i, \omega^{(i)}) A^-_{\alpha}(L+i, \omega^{(i)}) + \sum_{j=0}^i \varepsilon_j, \qquad
  i=0, 1, 2, \ldots,
\end{equation}
where
\begin{equation}
\begin{split}
  &A^{\pm}_{\alpha}(L+i,\omega^{(i)}) = \mp \sqrt{f} \frac{d}{dr} \sqrt{f} + W_{\alpha}(r;L+i,\omega^{(i)}), \\
  &\varepsilon_0 = E^{(\alpha)}_0(L,\omega), \\
  &\varepsilon_i = \alpha(4L+8i+2) + 2\Delta, \qquad i=1, 2, \ldots, 
\end{split}
\end{equation}
with
\begin{equation}
\begin{split}
  &W_{\alpha}(r;L+i,\omega^{(i)}) = - (L+i+1)\frac{f}{r} + \alpha \left(L + 2i + \frac{3}{2} + \frac{\Delta}{2\alpha}
      \right)r, \\
  &\omega^{(i)} = \sqrt{\omega^2 + +4i\alpha\Delta + 4i^2 \alpha^2},
\end{split}
\end{equation}
and $\omega^{(0)} = \omega$, $\omega^{(1)} = \omega'$. Such Hamiltonians are associated with a set of potentials $V_i(r)$, $i=0$, 1, 2, \ldots, in such a way that
\begin{equation}
  H_i = \pi_r^2 + V_i(r),
\end{equation}
where
\begin{equation}
  V_i(r) = V(r;L+i,\omega^{(i)}) + i\alpha\left(2L+2i+1+\frac{\Delta}{\alpha}\right).
\end{equation}
\par
%
%
One may also write
\begin{equation}
  H_{i+1} = A^-_{\alpha}(L+i,\omega^{(i)}) A^+_{\alpha}(L+i,\omega^{(i)}) + \sum_{j=0}^i \varepsilon_j.
\end{equation}
In other words, the first-order operators $A^{\pm}_{\alpha}(L+i,\omega^{(i)})$ fulfil a deformed shape invariance condition
\begin{align}
  &A^-_{\alpha}(L+i,\omega^{(i)}) A^+_{\alpha}(L+i,\omega^{(i)}) = A^+_{\alpha}(L+i+1,\omega^{(i+1)})
     A^-_{\alpha}(L+i+1,\omega^{(i+1)}) + \varepsilon_{i+1}, \nonumber \\
  &\qquad i=0, 1, 2, \ldots.
\end{align}
\par
%
%
{}Finally, one may note that the energy eigenvalues of the starting Hamiltonian $H_0$ can be directly found from the $\varepsilon_i$'s as
\begin{equation}
  E^{(\alpha)}_n(L,\omega) = \sum_{i=0}^n \varepsilon_i.
\end{equation}
\par
%
%
\section{\boldmath Rational extensions of the radial harmonic oscillator potential in the $m(\alpha;r)$ background}

\setcounter{equation}{0}

Rational extensions of the PT I potential have been constructed in the literature as partners of conventional potentials in connection with Jacobi EOP's. We plan to consider here more specifically the simplest ones obtained in one step SUSY \cite{cq08, cq09b, odake09}. By applying the inverse of the PCT presented in the previous section, we will build from them rational extensions of the radial harmonic oscillator potential in the $m(\alpha;r)$ PDM background and  study some properties of the latter.\par
%
%
\subsection{Rational extensions of the PT I potential}

The rational extensions of the PT I potential belong to three different types I, II, and III, according to the kind of seed function that is used to construct the partner. Such seed functions and their corresponding energies can be written as \cite{cq09b, bagchi15}
\begin{equation}
\begin{split}
  &\varphi^{\rm I}_m(u;A,B) = \phi_m(u; A,1-B), \qquad e^{\rm I}_m(A,B) = (A-B+1+2m)^2, \\
  & \quad B > m + \tfrac{1}{2}, \\
  &\varphi^{\rm II}_m(u;A,B) = \phi_m(u; 1-A,B), \qquad e^{\rm II}_m(A,B) = (B-A+1+2m)^2, \\
  & \quad A > m + \tfrac{1}{2}, \\
  &\varphi^{\rm III}_m(u;A,B) = \phi_m(u; 1-A,1-B), \qquad e^{\rm III}_m(A,B) = (-A-B+2+2m)^2,
       \\ 
  & \quad A, B > m + \tfrac{1}{2}, \qquad m \text{\ even}, 
\end{split}  \label{eq:varphi}
\end{equation}
where
\begin{equation}
  \phi_m(u;A,B) = (\sin u)^A (\cos u)^B P_m^{(A-\frac{1}{2}, B-\frac{1}{2})}(\cos 2u).  \label{eq:phi-m}
\end{equation}
\par
%
%
To obtain for the partner some rationally-extended PT I potential with given $A$ and $B$ parameters, we have to start from a conventional potential with different parameters $A'$ and $B'$, which depend on the type considered. The results read
\begin{equation}
\begin{split}
  U_0(u) &= U(u;A',B'), \\
  U_1(u) &= U(u;A',B') - 2 \frac{d^2}{du^2} \log \varphi_m(u;A',B') \\
  &= U^{(m)}_{\rm ext}(u;A,B) = U(u;A,B) + U^{(m)}_{\rm rat}(u;A,B),
\end{split}  \label{eq:U_0-U_1}
\end{equation}
where
\begin{equation}
  U^{(m)}_{\rm rat}(u;A,B) = 8\left\{z \frac{\dot{g}_m^{(A,B)}}{g_m^{(A,B)}} - (1-z^2)\left[
  \frac{\ddot{g}_m^{(A,B)}}{g_m^{(A,B)}} - \left(\frac{\dot{g}_m^{(A,B)}}{g_m^{(A,B)}}\right)^2\right]
  \right\}, \qquad z = \cos 2u.
\end{equation}
Here a dot denotes a derivative with respect to $z$, and we have for the three different types
\begin{equation}
\begin{split}
  \text{(I)\ } &A'=A-1, \quad B'=B+1, \quad g_m^{(A,B)}(z) = P_m^{\left(A-\frac{3}{2}, 
       -B-\frac{1}{2}\right)}(z), \quad B > m-\tfrac{1}{2}; \\
  \text{(II)\ } &A'=A+1, \quad B'=B-1, \quad g_m^{(A,B)}(z) = P_m^{\left(-A-\frac{1}{2}, 
       B-\frac{3}{2}\right)}(z), \quad A > m-\tfrac{1}{2};  \\  
  \text{(III)\ } &A'=A+1, \quad B'=B+1, \quad g_m^{(A,B)}(z) = P_m^{\left(-A-\frac{1}{2}, 
       -B-\frac{1}{2}\right)}(z), \quad A, B > m-\tfrac{1}{2}, \\
  & \quad m \text{\ even}.     
\end{split}  \label{eq:g}
\end{equation}
\par
%
%
{}For types I and II, the two partners $U_0(u)$ and $U_1(u)$ are strictly isospectral, so that the spectrum of the latter is given by
\begin{equation}
  \epsilon_{n}^{(\rm ext)}(A,B) = (A+B+2n)^2, \qquad n=0, 1, 2, \ldots, \qquad \text{for type I or II},
\end{equation}
whereas, for type III, the inverse of the seed function being normalizable, we get an extra bound state below the spectrum of $U_0(u)$, hence
\begin{equation}
  \epsilon_{n}^{(\rm ext)}(A,B) = (A+B+2n+2)^2, \qquad n = -m-1, 0, 1, 2, \ldots, \qquad \text{for type III}.
\end{equation}
\par
%
%
The wavefunctions of $U_1(u)$ corresponding to $n=0$, 1 2, \ldots\ can be obtained  from
\begin{equation}
  \phi^{(\rm ext}_n(u;A,B) \propto \frac{{\cal W}(\varphi^i_m(u;A',B'), \phi_n(u;A',B') \mid u)}{\varphi^i_m
  (u;A',B')}, \qquad i = \text{{\rm I}, {\rm II}, or {\rm III}}, 
\end{equation}
where ${\cal W}(\varphi^i_m, \phi_n \mid u)$ denotes the Wronskian of the functions $\varphi^i_m$ and $\phi_n$ \cite{muir}. From Eqs.~(\ref{eq:phi}), (\ref{eq:varphi}), (\ref{eq:g}), and standard properties of Wronskians, they can be rewritten as
\begin{equation}
  \phi^{\rm ext}_n(u;A,B) \propto \frac{\phi_0(u;A,B)}{g^{(A,B)}_m(z)} Q^{(m)}_n(z;A,B), \qquad n=0,1,2, \ldots,
  \label{eq:phi-ext}
\end{equation}
where $Q^{(m)}_n(z;A,B)$ is given by
\begin{align}
  Q^{(m)}_n(z;A,B) &= \left(B+\frac{1}{2}\right) g^{(A,B)}_m(z) P^{(A-\frac{3}{2},B+\frac{1}{2})}_n(z) 
       \nonumber \\
  &\quad +\frac{1}{2}(1+z) \biggl[(n+A+B) g^{(A,B)}_m(z) P^{(A-\frac{1}{2}, B+\frac{3}{2})}_{n-1}(z)
       \nonumber \\
  &\quad - (m+A-B-1) g^{(A+1,B-1)}_{m-1}(z) P^{(A-\frac{3}{2},B+\frac{1}{2})}_n(z)\biggr],
\end{align}
\begin{align}
  Q^{(m)}_n(z;A,B) &= - \left(A+\frac{1}{2}\right) g^{(A,B)}_m(z) P^{(A+\frac{1}{2}, B-\frac{3}{2})}_n(z)
      \nonumber \\
  &\quad + \frac{1}{2}(1-z) \biggl[(n+A+B) g^{(A,B)}_m(z) P^{(A+\frac{3}{2}, B-\frac{1}{2})}_{n-1}(z)
      \nonumber \\
  &\quad -(m-A+B-1) g^{(A-1,B+1)}_{m-1}(z) P^{(A+\frac{1}{2}, B-\frac{3}{2})}_n(z)\biggr],
\end{align}
\begin{align}
  Q^{(m)}_n(z;A,B) &= [B-A - (A+B+1)z] g^{(A,B)}_m(z) P^{(A+\frac{1}{2}, B+\frac{1}{2})}_n(z) \nonumber \\
  &\quad +\frac{1}{2}(1-z^2) \biggl[(n+A+B+2) g^{(A,B)}_m(z)  P^{(A+\frac{3}{2},B+\frac{3}{2})}_{n-1}(z)
       \nonumber \\
  &\quad - (m-A-B) g^{(A-1,B-1)}_{m-1}(z) P^{(A+\frac{1}{2},B+\frac{1}{2})}_n(z)\biggr],  \label{eq:Q}
\end{align}
for types I, II, and III, respectively. In the first two cases, it is a $(m+n)$th-degree polynomial in $z$, while for type III, it is a $(m+n+1)$th-degree polynomial in $z$. In the latter case, Eq.~(\ref{eq:phi-ext}) is also valid for $n=-m-1$ and $Q^{(m)}_{-m-1}(z;A,B) = 1$.\par
%
%
Due to the orthogonality properties of bound-state wavefunctions, in all three cases, the polynomials $Q^{(m)}_n(u;A,B)$ constitute families of orthogonal polynomials on $(-1,+1)$ with respect to the measure $(1-z)^{A-\frac{1}{2}} (1+z)^{B-\frac{1}{2}} \left(g^{(A,B)}_m(z)\right)^{-2} dz$. From the absence of scattering states, it results that these families also form complete sets and therefore qualify as EOP families.\par
%
%
\subsection{\boldmath Rational extensions of the radial harmonic oscillator potential in the $m(\alpha;r)$ background}

On applying Eqs.~(\ref{eq:u-r}), (\ref{eq:AB}), and (\ref{eq:phi-m}), the seed functions (\ref{eq:varphi}) are transformed into
\begin{equation}
\begin{split}
  &\chi^{\rm I}_m(r;L,\omega) = r^{L+1} f^{-\frac{1}{2}(L+\frac{5}{2}-\frac{\Delta}{2\alpha})}
      P^{(L+\frac{1}{2}, -\frac{\Delta}{2\alpha})}_m(t), \qquad m < \frac{\Delta}{2\alpha},\\
  &\chi^{\rm II}_m(r;L,\omega) = r^{-L} f^{-\frac{1}{2}(-L+\frac{3}{2}+\frac{\Delta}{2\alpha})}
      P^{(-L-\frac{1}{2}, \frac{\Delta}{2\alpha})}_m(t), \qquad m < L+\frac{1}{2}, \\
  &\chi^{\rm III}_m(r;L,\omega) = r^{-L} f^{-\frac{1}{2}(-L+\frac{3}{2}-\frac{\Delta}{2\alpha})}
      P^{(-L-\frac{1}{2}, -\frac{\Delta}{2\alpha})}_m(t), \qquad m < L+\frac{1}{2}, \\
  &\qquad m < \frac{\Delta}{2\alpha}, \qquad \text{$m$ even},
\end{split} \label{eq:chi}
\end{equation}
with corresponding energies
\begin{equation}
\begin{split}
  &{\cal E}^{\rm I}_m(L,\omega) = \alpha \left(2L+ \frac{5}{2}\right) - \left(L+\frac{3}{2}\right)\Delta
      + 4\left[\alpha\left(L+\frac{3}{2}\right) - \frac{\Delta}{2}\right]m + 4\alpha m^2, \\
  &{\cal E}^{\rm II}_m(L,\omega) = \alpha \left(-2L+\frac{1}{2}\right) - \left(L-\frac{1}{2}\right) \Delta
     + 4\left[\alpha\left(-L+\frac{1}{2}\right) + \frac{\Delta}{2}\right] m + 4\alpha m^2, \\
  &{\cal E}^{\rm III}_m(L,\omega) = \alpha \left(-2L+\frac{1}{2}\right) + \left(L-\frac{1}{2}\right) \Delta
     + 4\left[\alpha\left(-L+\frac{1}{2}\right) - \frac{\Delta}{2}\right] m + 4\alpha m^2.
\end{split}
\end{equation}
As it can be checked, with the conditions on $m$ given in (\ref{eq:chi}), the seed functions $\chi^{\rm I}_m$, $\chi^{\rm II}_m$, and $\chi^{\rm III}_m$ are in the disconjugacy sector of the radial harmonic oscillator, i.e., ${\cal E}^{\rm I}_m(L,\omega) < E^{(\alpha)}_0(L,\omega)$, ${\cal E}^{\rm II}_m(L,\omega) < E^{(\alpha}_0(L,\omega)$, and ${\cal E}^{\rm III}_m(L,\omega) < E^{(\alpha)}_0(L,\omega)$.\par
%
%
{}Furthermore, from Eqs.~(\ref{eq:PCT3}), (\ref{eq:ab}), (\ref{eq:AB}), and (\ref{eq:c}), the partners $U_0(u)$ and $U_1(u)$, defined in Eqs.~(\ref{eq:U_0-U_1})--(\ref{eq:g}), give rise to partners
\begin{equation}
\begin{split}
  V_0(r) &= V(r;L',\omega '), \\
  V_1(r) &= V^{(m)}_{\rm ext}(r;L,\omega) + \gamma = V(r;L,\omega) + V^{(m)}_{\rm rat}(r;L,\omega) + \gamma,  
  \label{eq:V_0-V_1}
\end{split}
\end{equation}
where
\begin{equation}
  V^{(m)}_{\rm rat}(r;L,\omega) = 8\alpha \left\{t \frac{\dot{p}_m^{(L,\Delta)}}{p_m^{(L,\Delta)}}
  - (1-t^2) \left[\frac{\ddot{p}_m^{(L,\Delta)}}{p_m^{(L,\Delta)}} - \left(\frac{\dot{p}_m^{(L,\Delta)}}
  {p_m^{(L,\Delta)}}\right)^2\right]\right\} \label{eq:Vrat1}
\end{equation}
and a dot denotes a derivative with respect to $t = \frac{1-\alpha r^2}{1+\alpha r^2}$. Here, for the three different types, we get
\begin{equation}
\begin{split}
  \text{(I)\ } &L'=L-1, \quad \omega' = \sqrt{\omega^2+4\alpha^2+4\alpha\Delta}, \quad p_m^{(L,\Delta)}(t)
      = P_m^{(L-\frac{1}{2}, -\frac{\Delta}{2\alpha}-1)}(t), \\
  &\gamma = \alpha(2L-1) - \Delta, \quad m< \frac{\Delta}{2\alpha}+1, \\
  \text{(II)\ }&L'=L+1, \quad \omega' = \sqrt{\omega^2+4\alpha^2-4\alpha\Delta}, \quad p_m^{(L,\Delta)}(t)
      = P_m^{(-L-\frac{3}{2}, \frac{\Delta}{2\alpha}-1)}(t), \\
  &\gamma = -\alpha(2L+3)+\Delta, \quad m<L+\frac{3}{2}, \quad \alpha < \frac{\omega}{2\sqrt{2}}, \\
  \text{(III)\ }&L'=L+1, \quad \omega' = \sqrt{\omega^2+4\alpha^2+4\alpha\Delta}, \quad p_m^{(L,\Delta)}(t)
      = P_m^{(-L-\frac{3}{2}, -\frac{\Delta}{2\alpha}-1)}(t), \\
  &\gamma = -\alpha(2L+3) -\Delta, \quad m<L+\frac{3}{2}, \quad m<\frac{\Delta}{2\alpha}+1, \quad \text{$m$ even}. 
\end{split}. \label{eq:Vrat2}
\end{equation}
Note that the presence of the additive constant $\gamma$ in (\ref{eq:V_0-V_1}) is due to the dependence of $c$ on $L$ and $\omega$ (see Eq.~(\ref{eq:c})) and that the latter assume different values for the two partners. So for type I, for instance, the PCT changes $U(u;A-1,B+1)$ into $V(r;L-1,\omega') = \alpha U(u;A-1,B+1)-\alpha(L-1)L-\frac{\omega^{\prime2}}{4\alpha}$, whereas $U^{(m)}_{\rm ext}(u;A,B)$ is modified into $V^{(m)}_{\rm ext}(r;L,\omega) = \alpha U^{(m)}_{\rm ext}(u;A,B) - \alpha L(L+1) - \frac{\omega^2}{4\alpha}$. Since $U(u;A-1,B+1)$ and $U^{(m)}_{\rm ext}(u;A,B)$ are isospectral, to get the same property for the image potentials, we have to consider $V(r;L-1,\omega')$ and $V^{(m)}_{\rm ext}(r;L,\omega) + \alpha L(L+1) + \frac{\omega^2}{4\alpha} - \alpha L(L-1) - \frac{\omega^{\prime2}}{4\alpha} = V^{(m)}_{\rm ext}(r;L,\omega) + \alpha (2L-1) - \Delta$. A similar reasoning applies to the remaining two types.\par
%
%
{}For type I or II, the spectra of $V_0(r)$ and $V_1(r)$ are given by $E^{(\alpha)}_n(L',\omega')$, $n=0$, 1, 2,\ldots. Hence, the spectrum of $V^{(m)}_{\rm ext}(r;L,\omega)$, given by $E^{(\rm ext)}_n(L,\omega) = E^{(\alpha)}_n(L',\omega') - \gamma$, turns out to be the same as that of $V(r;L,\omega)$, i.e.,
\begin{align}
  E^{(\rm ext)}_n(L,\omega) &= \alpha\left(2L+\frac{5}{2}\right) + \left(L+\frac{3}{2}\right)\Delta + 4\left[\alpha
      \left(L+\frac{3}{2}\right) + \frac{\Delta}{2}\right]n + 4\alpha n^2, \nonumber \\
  &\quad n=0,1,2, \ldots,\quad \text{for type I or II}.
\end{align}
On the other hand, we get
\begin{align}
  E^{(\rm ext)}_n(L,\omega) &= \alpha \left(6L+\frac{25}{2}\right) + \left(L+\frac{7}{2}\right)\Delta + 4\left[\alpha
      \left(L+\frac{7}{2}\right) + \frac{\Delta}{2}\right]n + 4\alpha n^2, \nonumber \\
  &\quad n = -m-1, 0,1,2, \ldots,\quad \text{for type III}.
\end{align}
\par
%
%
The corresponding wavefunctions, obtained from Eqs.~(\ref{eq:psi-phi}) and (\ref{eq:phi-ext})--(\ref{eq:Q}), read
\begin{equation}
  \psi_n^{(\rm ext)}(r;L,\omega) \propto \frac{\psi^{(\alpha)}_0(r;L,\omega)}{p_m^{(L,\Delta)}(t)} 
  Q^{(m)}_n(t;L,\omega), \qquad t = \frac{1-\alpha r^2}{1+\alpha r^2}, \label{eq:wf-ext-1}
\end{equation}
where the polynomials $Q^{(m)}_n(t;L,\omega)$ can be written as
\begin{align}
  &Q^{(m)}_n(t;L,\omega) = \left(\frac{\Delta}{2\alpha}+1\right) p_m^{(L,\Delta)}(t) P_n^{(L-\frac{1}{2},1+\frac{\Delta}
       {2\alpha})}(t) \nonumber \\
  &\quad + \frac{1}{2}(1+t) \biggl[\biggl(n+L+\frac{3}{2}+\frac{\Delta}{2\alpha}\biggr) p_m^{(L,\Delta)}(t) 
        P_{n-1}^{(L+\frac{1}{2},2+\frac{\Delta}{2\alpha})}(t) \nonumber \\
  &\quad - \left(m+L-\frac{1}{2}-\frac{\Delta}{2\alpha}\right) p_{m-1}^{(L+1,\Delta-2\alpha)}(t) P_n^{(L-\frac{1}{2},
       1+\frac{\Delta}{2\alpha})}(t)\biggr], \qquad n=0,1,2,\ldots,  \label{eq:wf-ext-1a}
\end{align}
\begin{align}
  &Q^{(m)}_n(t;L,\omega) = -\left(L+\frac{3}{2}\right) p_m^{(L,\Delta)}(t) P_n^{(L+\frac{3}{2},-1+\frac{\Delta}{2\alpha})}(t)
        \nonumber \\
  &\quad+\frac{1}{2}(1-t) \biggl[\left(n+L+\frac{3}{2}+\frac{\Delta}{2\alpha}\right) p_m^{(L,\Delta)}(t) P_{n-1}^{(L+\frac{5}
        {2},\frac{\Delta}{2\alpha})}(t) \nonumber \\
  &\quad-\left(m-L-\frac{3}{2}+\frac{\Delta}{2\alpha}\right) p_{m-1}^{(L-1,\Delta+2\alpha)}(t) P_n^{(L+\frac{3}{2},-1
        +\frac{\Delta}{2\alpha})}(t)\biggr], \qquad n=0,1,2,\ldots,
\end{align}
and
\begin{align}
  &Q^{(m)}_n(t;L,\omega) = - \left[L+\frac{1}{2}-\frac{\Delta}{2\alpha} + \left(L+\frac{5}{2}+\frac{\Delta}{2\alpha}\right)t
       \right] p_m^{(L,\Delta)}(t) P_n^{(L+\frac{3}{2},1+\frac{\Delta}{2\alpha})}(t) \nonumber \\
  &\quad+ \frac{1}{2}(1-t^2) \biggl[\left(n+L+\frac{7}{2}+\frac{\Delta}{2\alpha}\right) p_m^{(L,\Delta)}(t) P_{n-1}
       ^{(L+\frac{5}{2},2+\frac{\Delta}{2\alpha})}(t) \nonumber \\
  &\quad-\left(m-L-\frac{3}{2}-\frac{\Delta}{2\alpha}\right) p_{m-1}^{(L-1,\Delta-2\alpha)}(t) P_n^{(L+\frac{3}{2},
       1+\frac{\Delta}{2\alpha})}(t)\biggr], \qquad n=0,1,2,\ldots, \nonumber\\
  &Q^{(m)}_{-m-1}(t;L,\omega) = 1,  \label{eq:wf-ext-2}
\end{align}
for types I, II, and III, respectively. They constitute orthogonal and complete families of polynomials (i.e., EOP's) on $(-1,+1)$ with respect to the measure $(1-t)^{L+\frac{1}{2}} (1+t)^{\frac{\Delta}{2\alpha}} \left(p_m^{(L,\Delta)}(t)\right)^{-2}dt$.\par
%
%
When $\alpha$ goes to zeroi so that Eq.~(\ref{eq:def-SE}) gives back the constant-mass Schr\"odinger equation (\ref{eq:SE}), its extensions obtained by replacing $V(r;L,\omega)$ by $V_m^{(\rm ext)}(r;L,\omega)$ give rise to known results for the constant-mass rationally-extended radial harmonic oscillator potential, associated with Laguerre EOP's of type I, II, or III \cite{cq11, marquette}.\par
%
%
On extensively using Eqs.~(\ref{eq:limit-1}) and (\ref{eq:limit-2}), it is indeed easy to show that the seed functions (\ref{eq:chi}) become
\begin{equation}
\begin{split}
  &\lim_{\alpha\to 0} \chi^{\rm I}_m(r;L,\omega) = r^{L+1} e^{\frac{1}{4}\omega r^2} L_m^{(L+\frac{1}{2})}\left(-\tfrac{1}{2}
       \omega r^2\right), \\
  &\lim_{\alpha\to 0} \chi^{\rm II}_m(r;L,\omega) = r^{-L} e^{-\frac{1}{4}\omega r^2} L_m^{(-L-\frac{1}{2})}\left(\tfrac{1}{2}
       \omega r^2\right), \qquad m<L+\frac{1}{2}, \\
  &\lim_{\alpha\to 0} \chi^{\rm III}_m(r;L,\omega) = r^{-L} e^{\frac{1}{4}\omega r^2} L_m^{(-L-\frac{1}{2})}\left(-\tfrac{1}{2}
       \omega r^2\right), \qquad m < L+\frac{1}{2}, \qquad \text{$m$ even},
\end{split}
\end{equation}
with corresponding energies
\begin{equation}
\begin{split}
  &\lim_{\alpha\to 0} {\cal E}^{\rm I}_m(L,\omega) = - \omega\left(L + \tfrac{3}{2} + 2m\right), \\
  &\lim_{\alpha\to 0} {\cal E}^{\rm II}_m(L,\omega) = - \omega\left(L - \tfrac{1}{2} - 2m\right), \\
  &\lim_{\alpha\to 0} {\cal E}^{\rm III}_m(L,\omega) = - \omega\left(- L + \tfrac{1}{2} + 2m\right).
\end{split}
\end{equation}
Furthermore, to compare the extended potentials, defined in Eqs.~(\ref{eq:V_0-V_1})--(\ref{eq:Vrat2}), with known results for the constant-mass rationally-extended radial harmonic oscillator potential, we have first to convert derivatives with respect to $t$ into derivatives with respect to $r$, then to use Eq.~(\ref{eq:limit-2}), and finally to change derivatives with respect to $r$ into derivatives with respect to $\rho = \frac{1}{2}\omega r^2$ (denoted by a hat). The result reads
\begin{equation}
  \lim_{\alpha\to 0} V^{(m)}_{\rm rat}(r;L,\omega) = - 2\omega\left\{\frac{\hat{q}^{(L)}_m}{q^{(L)}_m} + 2\rho \left[
  \frac{\hat{\hat{q}}^{(L)}_m}{q^{(L)}_m} - \left(\frac{\hat{q}^{(L)}_m}{q^{(L)}_m}\right)^2\right]\right\},
\end{equation}
where
\begin{equation}
  q^{(L)}_m(\rho) = \begin{cases}
     L^{(L-\frac{1}{2})}_m(-\rho) & \text{for type I}, \\
     L^{(-L-\frac{3}{2})}_m(\rho) & \text{for type II}, \\
     L^{(-L-\frac{3}{2})}_m(-\rho) & \text{for type III},
\end{cases}
\end{equation}
in agreement with Refs.~\cite{cq11,marquette}.\par
%
%
{}Finally, considering the extended potential wavefunctions, given in Eqs.~(\ref{eq:wf-ext-1})--(\ref{eq:wf-ext-2}), from Eqs.~(\ref{eq:limit-1}) and (\ref{eq:limit-2}) we get the following results:
\begin{equation}
  \lim_{\alpha\to 0} \psi^{(\rm ext)}_n(r;L,\omega) \propto \frac{r^{L+1}e^{-\frac{1}{4}\omega r^2}}{q^{(L)}_m(\rho)}
  \lim_{\alpha\to 0} Q^{(m)}_n(t;L,\omega),
\end{equation}
where
\begin{equation}
\begin{split}
  &\lim_{\alpha\to 0} Q^{(m)}_n(t;L,\omega) \\
  &\quad = \begin{cases}
    q^{(L)}_m(\rho) \left[L^{(L-\frac{1}{2})}_n(\rho) + L^{(L+\frac{1}{2})}_{n-1}(\rho)\right] 
       + q^{(L+1)}_{m-1}(\rho)L^{(L-\frac{1}{2})}_n(\rho) & \\[0.2cm]
    \quad\text{for type I}, &, \\[0.2cm]
    q^{(L)}_m(\rho) \left[(L+\frac{3}{2})L^{(L+\frac{3}{2})}_n(\rho) - \rho L^{(L+\frac{5}{2})}_{n-1}(\rho)\right]
       + \rho q^{(L-1)}_{m-1}(\rho) L^{(L+\frac{3}{2})}_n(\rho)  & \\[0.2cm]
    \quad\text{for type II}, &, \\[0.2cm]
    q^{(L)}_m(\rho) \left[(L+\frac{3}{2}-\rho)L^{(L+\frac{3}{2})}_n(\rho) - \rho L^{(L+\frac{5}{2})}_{n-1}(\rho)\right]
       - \rho q^{(L-1)}_{m-1}(\rho) L^{(L+\frac{3}{2})}_n(\rho)  & \\[0.2cm]
    \quad\text{for type III},&
  \end{cases}
\end{split}
\end{equation}
with $n=0$, 1, 2, \ldots, and $\lim_{\alpha\to 0} Q^{(m)}_{-m-1}(t;L,\omega) = 1$ in the type III case. This is again consistent with the results of Refs.~\cite{cq11,marquette}.\par
%
%
\subsection{Deformed supersymmetry properties}

By the inverse of the PCT defined in Sec.~2.1, the conventional SUSY relation between the partners $U_0(u)$ and $U_1(u)$ of Eq.~(\ref{eq:U_0-U_1}) is converted into a deformed SUSY relation between the partners $V_0(r)$ and $V_1(r)$ of Eq.~(\ref{eq:V_0-V_1}). The superpotential characterizing the latter is given by
\begin{equation}
  W^{(m)}_{\alpha}(r;L',\omega') = - f \frac{d}{dr} \log\chi_m(r;L',\omega') - \frac{1}{2} f',
\end{equation}
which yields
\begin{equation}
\begin{split}
  &W^{(m)}_{\alpha}(r;L-1, \sqrt{\omega^2+\alpha^2+4\alpha\Delta}) = - \frac{L}{r} - \alpha r\left(\frac{1}{2}
     + \frac{\Delta}{2\alpha}\right) - f \frac{p'_m}{p_m} \quad \text{for type I}, \\
  &W^{(m)}_{\alpha}(r;L+1, \sqrt{\omega^2+\alpha^2-4\alpha\Delta}) = \frac{L+1}{r} - \alpha r\left(\frac{1}{2}
     - \frac{\Delta}{2\alpha}\right) - f \frac{p'_m}{p_m} \quad \text{for type II}, \\
  &W^{(m)}_{\alpha}(r;L+1, \sqrt{\omega^2+\alpha^2+4\alpha\Delta}) = \frac{L+1}{r} - \alpha r\left(\frac{1}{2}
     + \frac{\Delta}{2\alpha}\right) - f \frac{p'_m}{p_m} \quad \text{for type III},
\end{split}
\end{equation}
the corresponding $p_m = p_m^{(L,\Delta)}(t)$ being defined in (\ref{eq:Vrat2}) and a dash denoting as before a derivative with respect to $r$.\par
%
%
As it can be checked by using the differential equation satisfied by Jacobi polynomials \cite{olver}, in the three cases we can write
\begin{equation}
\begin{split}
  &\left[W^{(m)}_{\alpha}(r;L',\omega')\right]^2 - f W^{(m)\prime}_{\alpha}(r;L',\omega') + {\cal E}_m(L',\omega')
      = V(r;L',\omega'), \\
  &\left[W^{(m)}_{\alpha}(r;L',\omega')\right]^2 + f W^{(m)\prime}_{\alpha}(r;L',\omega') + {\cal E}_m(L',\omega')
      = V^{(m)}_{\rm ext}(r;L,\omega) + \gamma.
\end{split}
\end{equation}
\par
%
%
As shown in Sec.~2.2, the starting potential $V(r;L',\omega')$ in this deformed SUSY construction satisfies a deformed shape invariance property with a partner given by $V(r;L'+1,\omega'') + \alpha(2L'+3+ \frac{\Delta'}{\alpha})$, where $\Delta' = \sqrt{\omega^{\prime 2}+ \alpha^2}$ and $\omega'' = \sqrt{\omega^{\prime2}+4\alpha^2+4\alpha\Delta'}$. One may wonder whether the final potential $V^{(m)}_{\rm ext}(r;L,\omega) + \gamma$ in such a construction has a similar property. As we plan to show, the answer is affirmative in the two isospectral cases I and II.\par
%
%
Let us indeed now consider a superpotential
\begin{equation}
  W^{(m)}_{\rm ext}(r;L,\omega) = - f \frac{d}{dr} \log \psi^{(\rm ext)}_0(r;L,\omega) - \frac{1}{2}f' \label{eq:W-m-ext}
\end{equation}
with
\begin{equation}
  \psi^{(\rm ext)}_0(r;L,\omega) \propto \frac{\psi^{(\alpha)}_0(r;L,\omega)}{p_m^{(L,\Delta)}(t)} Q^{(m)}_0(t;L,
  \omega),  \label{eq:psi-ext-0}
\end{equation}
and
\begin{equation}
  Q^{(m)}_0(t;L,\omega) = \left(\frac{\Delta}{2\alpha}+1\right) p_m^{(L,\Delta)}(t) - \frac{1}{2}\left(m+L-\frac{1}{2}
  - \frac{\Delta}{2\alpha}\right) (1+t) p_{m-1}^{(L+1,\Delta-2\alpha)}(t)
\end{equation}
or
\begin{equation}
  Q^{(m)}_0(t;L,\omega) = - \left(L+\frac{3}{2}\right) p_m^{(L,\Delta)}(t) - \frac{1}{2}\left(m-L-\frac{3}{2}
  + \frac{\Delta}{2\alpha}\right) (1-t) p_{m-1}^{(L-1,\Delta+2\alpha)}(t)
\end{equation}
for type I or II, respectively. On using the definition of $p_m^{(L,\Delta)}(t)$, given in Eq.~(\ref{eq:Vrat2}), and the expansion of these polynomials into powers \cite{olver}, it is straightforward to rewrite $Q^{(m)}_0(t;L,\omega)$ as
\begin{equation}
  Q^{(m)}_0(t;L,\omega) = \left(1 + \frac{\Delta}{2\alpha} - m\right) p_m^{(L+1,\Delta+2\alpha)}(t) \qquad \text{for type I}
  \label{eq:Q-I}
\end{equation}
and
\begin{equation}
  Q^{(m)}_0(t;L,\omega) = \left(m -L - \frac{3}{2}\right) p_m^{(L+1,\Delta+2\alpha)}(t) \qquad \text{for type II}.
  \label{eq:Q-II}
\end{equation}
Combining Eqs.~(\ref{eq:W-m-ext}) and (\ref{eq:psi-ext-0}) with (\ref{eq:Q-I}) or (\ref{eq:Q-II}) yields
\begin{equation}
  W^{(m)}_{\rm ext}(r;L,\omega) = - \frac{L+1}{r} + \alpha\left(\frac{1}{2} + \frac{\Delta}{2\alpha}\right) r - f
  \left(\frac{p_m^{(L+1,\Delta+2\alpha)\prime}}{p_m^{(L+1,\Delta+2\alpha)}} 
  - \frac{p_m^{(L,\Delta)\prime}}{p_m^{(L,\Delta)}}\right)
\end{equation}
for both types. It is then straightforward to show that the partner of $V^{(m)}_{\rm ext}(r;L,\omega)$ is given by
\begin{equation}
  V^{(m)}_{\rm ext}(r;L,\omega) + 2f W^{(m)\prime}_{\rm ext}(r;L,\omega) = V^{(m)}_{\rm ext}(r;L+1,\omega')
  + \alpha\left(2L+3 + \frac{\Delta}{\alpha}\right),
\end{equation}
which proves the deformed shape invariance property of type I and II extended potentials.\par
%
%
\subsection{Explicit example}

The simplest example of rationally-extended radial harmonic oscillator associated with the PDM $m(\alpha;r)$ corresponds to $m=1$ in (\ref{eq:V_0-V_1}) and (\ref{eq:Vrat1}) and may be obtained as a type I or II potential. Considering for instance the former case, from
\begin{align}
  p_1^{(L,\Delta)}(t) &= P_1^{(L-\frac{1}{2}, -\frac{\Delta}{2\alpha}-1)}(t) \nonumber \\
  &= \frac{1}{2}\left[\left(L - \frac{\Delta}{2\alpha} + \frac{1}{2} \right) t + L + \frac{\Delta}{2\alpha} + \frac{1}{2}
       \right] \nonumber \\
  &= \frac{\Delta r^2 + 2L +1}{2(1+\alpha r^2)},
\end{align}
we get the extended potential
\begin{equation}
  V^{(1)}_{\rm ext}(r;L,\omega) = \frac{L(L+1)}{r^2} + \frac{1}{4} \omega^2 r^2 + 4[\Delta - \alpha(2L+1)]
  \frac{(1+\alpha r^2)(\Delta r^2-2L-1)}{(\Delta r^2+2L+1)^2}.  \label{eq:pot-1}
\end{equation}
\par
%
%
Its spectrum remains given by Eq.~(\ref{eq:E}). Its wavefunctions, expressed in Eqs.~(\ref{eq:wf-ext-1}) and (\ref{eq:wf-ext-1a}), may also be written in a simpler way in terms of the $(n+1)$th-degree X$_1$-Jacobi EOP's $\hat{P}_{n+1}^{(L+\frac{1}{2}, \frac{\Delta}{2\alpha})}(t)$, $n=0$, 1, 2, \ldots, defined in Ref.~\cite{gomez09}. This is done by applying the PCT considered in Sec.~2.2 on a known result for a rationally-extended PT I potential in a constant-mass background \cite{cq09b}. The result reads
\begin{equation}
  \psi_n^{(\rm ext)}(r;L,\omega) = {\cal N}^{(\rm ext)}_{n,L,\omega} \frac{r^{L+1} f^{-\frac{1}{2}(L+\frac{1}{2}+\frac{
  \Delta}{2\alpha})}}{2L+1+\Delta r^2} \hat{P}_{n+1}^{(L+\frac{1}{2},\frac{\Delta}{2\alpha})}(t), \qquad n=0, 1, 2, \ldots,  \label{eq:psi-ext-1}
\end{equation}. 
with $t = \frac{1-\alpha r^2}{1+\alpha r^2}$ and 
\begin{equation}
  {\cal N}^{(\rm ext)}_{n,L,\omega} = \left(\frac{8 \alpha^{L+\frac{3}{2}} \left(L+\frac{1}{2}-\frac{\Delta}{2\alpha}\right)^2
  \left(L+\frac{3}{2}+\frac{\Delta}{2\alpha}+2n\right) n! \Gamma\left(L+\frac{3}{2}+\frac{\Delta}{2\alpha}+n\right)}
  {\left(L+\frac{3}{2}+n\right) \left(n+1+\frac{\Delta}{2\alpha}\right) \Gamma\left(L+\frac{1}{2}+n\right)
  \Gamma\left(n+\frac{\Delta}{2\alpha}\right)}\right)^{1/2}.  \label{eq:psi-ext-2}
\end{equation}
For the lowest $n$ values, for instance, we obtain
\begin{equation}
\begin{split}
  \hat{P}_1^{(L+\frac{1}{2},\frac{\Delta}{2\alpha})}(t) &= - \frac{1}{2\left(L+\frac{1}{2}-\frac{\Delta}{2\alpha}\right)
      (1+\alpha r^2)} [2L+3 + (\Delta+2\alpha) r^2], \\
  \hat{P}_2^{(L+\frac{1}{2},\frac{\Delta}{2\alpha})}(t) &= - \frac{1}{4\left(L+\frac{1}{2}-\frac{\Delta}{2\alpha}\right)
      (1+\alpha r^2)^2} [(2L+1)(2L+5) - \Delta(\Delta+4\alpha) r^4], \\
  \hat{P}_3^{(L+\frac{1}{2},\frac{\Delta}{2\alpha})}(t) &= - \frac{1}{16\left(L+\frac{1}{2}-\frac{\Delta}{2\alpha}\right)
      (1+\alpha r^2)^3} [(2L+1)(2L+3)(2L+7) \\
  &\quad -(2L+1)(2L+7)(\Delta+6\alpha) r^2 - (2L+7)\Delta(\Delta+6\alpha) r^4 \\
  &\quad + \Delta(\Delta+2\alpha)(\Delta+6\alpha) r^6].  \label{eq:psi-ext-3}
\end{split}
\end{equation}
%
%
In Fig.~3, we show potential (\ref{eq:pot-1}) for $L=\omega=1$ and several values of the deforming parameter $\alpha$. The first three wavefunctions of this potential are displayed for $\alpha=1/\sqrt{3}$ in Fig.~4. The corresponding eigenvalues are $E_0^{(1/\sqrt{3})}(1,1)=\frac{19}{2\sqrt{3}}$, $E_1^{(1/\sqrt{3})}(1,1)=\frac{55}{2\sqrt{3}}$, and $E_2^{(1/\sqrt{3})}(1,1)=\frac{107}{2\sqrt{3}}$.\par
%
%
\begin{figure}
\begin{center}
\includegraphics{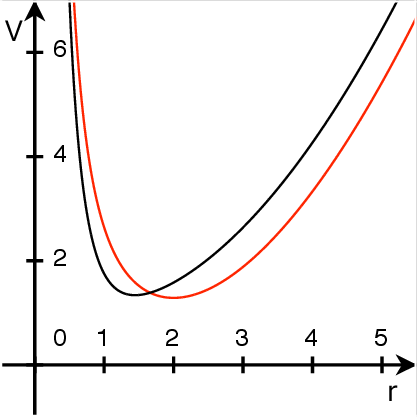}
\caption{Plots of the extended potential $V^{(1)}_{\rm ext}(r;L,\omega)$ in terms of $r$ for $\alpha=1/\sqrt{3}$ (red line) and $\alpha=0$ (black line). The parameter values are $L=\omega=1$.}
\end{center}
\end{figure}
\par
%
%
\begin{figure}
\begin{center}
\includegraphics{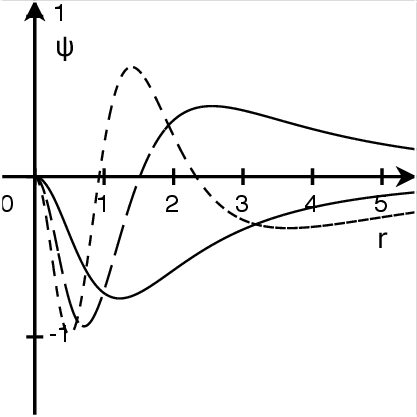}
\caption{Plots of the wavefunction $\psi^{(\rm ext)}_n(r;L,\omega)$, defined in Eqs.~(\ref{eq:psi-ext-1})--(\ref{eq:psi-ext-3}), in terms of $r$ for $n=0$ (full line), $n=1$ (long-dashed line), and $n=2$ (short-dashed line). The parameter values are $L=\omega=1$ and $\alpha=1/\sqrt{3}$.}
\end{center}
\end{figure}
\par
%
%
\section{Conclusion}

In the present paper, we have shown that the problem of the radial harmonic oscillator in a $m(\alpha;r) = (1+\alpha r^2)^{-2}$ background can be easily solved by mapping the corresponding Schr\"odinger equation onto that for the PT I potential with constant mass by means of a PCT. In addition, we have proved that the well-known shape invariance of the PT I potential in SUSYQM gives rise to a deformed shape invariance property for the considered PDM problem in deformed SUSYQM.\par
%
%
We have then taken advantage of the knowledge of rational extensions of the PT I potential connected with $X_m$-Jacobi EOP's of type I, II, and III to transform them into rational extensions of the radial harmonic oscillator in the $m(\alpha;r)$ background. These results have been analyzed in the  deformed SUSYQM framework and the extended radial harmonic oscillator potentials of type I and II have been shown to be endowed with a deformed shape invariance property.\par
%
%
{}Finally, the spectrum and wavefunctions of the radial harmonic oscillator and its extensions in the $m(\alpha;r)$ background have been proved to go over for $\alpha\to 0$ to those corresponding to a constant mass, in which case Jacobi polynomials and $X_m$-Jacobi EOP's become Laguerre polynomials and $X_m$-Laguerre EOP's, respectively.\par
%
%
Considering multi-indexed rational extensions and corresponding orthogonal polynomials would be a very interesting topic for future investigation. Another open question for future work would be the possibility of transferring to the radial harmonic oscillator problem with PDM the more general one- and $m$-step extensions of the PT I potential based on the use of para-Jacobi polynomials \cite{bagchi15, grandati}.\par
%
%
\section*{Credit authorship contribution statement}

C.\ Quesne: Conceptualization, Methodology, Investigation, Writing - review and editing.\par
%
%
\section*{Declaration of competing interest}

The authors declare that they have no known competing financial interests or personal relationships that could have appeared to influence the work reported in this paper.\par
%
%
\section*{Data availability}

No data was used for the research described in the article.\par
%
%
\section*{Acknowledgment}

The author was supported by the Fonds de la Recherche Scientifique - FNRS under Grant Number 4.45.10.08.\par
%
%

\end{document}